\documentstyle[prl,aps,epsfig,twocolumn]{revtex}   

\begin{document}

\draft

\title{Atom Optics with Microfabricated Optical Elements}

\author{G. Birkl, F.B.J. Buchkremer, R. Dumke, and W. Ertmer}

\address{Institut f\"ur Quantenoptik, Universit\"at Hannover,
Welfengarten 1, 30167 Hannover}

\date{\today}
\maketitle

\begin{abstract} 

We introduce a new direction in the field of atom optics, atom interferometry, and 
neutral-atom quantum
information processing. It is based on
the use of microfabricated optical elements.
With these elements versatile 
and integrated atom optical devices can be created in a compact fashion.
This approach opens the possibility to scale, 
parallelize, and miniaturize
atom optics for new investigations in fundamental research and application. It will
lead to new, compact sources of ultracold atoms, compact sensors based on matter 
wave interference and new
approaches towards quantum computing with neutral atoms. 
The exploitation of the unique features of the quantum mechanical behavior of matter waves 
and the capabilities of powerful state-of-the-art micro- and nanofabrication techniques
lend this approach a special attraction.

\end{abstract}

\narrowtext


\section {Introduction}

The investigation and exploitation of the wave properties of atomic matter is of great 
interest for fundamental as well as applied research and therefore constitutes one of the
most active areas in atomic physics and quantum optics. %
Profiting from the enormous progress in laser cooling 
\cite{Lasercooling},
the field of atom optics has been established and research in this area has already 
led to many exciting results:
Various types of atom optical elements (lenses, mirrors, beam splitters, etc.) 
\cite{Atomoptics} and
atom interferometers 
%
%
\cite{Atominterferometry} have been realized and precise atom interferometrical
measurements of fundamental constants \cite{Young}, atomic properties 
\cite{Chapman,Sterr,Gibble,Clairon}, 
acceleration forces \cite{Peters,Snadden} and rotations \cite{Riehle,Lenef,Gustavson} have been 
performed. 
With the development of 
reliable sources for coherent matter waves such as Bose-Einstein condensates 
\cite{BEC-Klassiker} and
laser-like atom beams \cite{Atomlaser} this work is in the process of being 
extended towards improved measurement schemes based on the properties of 
many-particle wavefunctions \cite{Bouyer_BECinterferometer}. 

As a consequence of all these developments, there is now a
vast interest in compact and reliable atom optical setups 
which not only expand the applicability of atom optics in fundamental
research, but also allow the technological implementation of
atom optical measurement systems. 
A new approach to this challenge lies in the development of 
miniaturized and integrated atom optical setups based on 
microfabricated structures. %

A number of groups have employed microfabricated {\bf mechanical} structures
%
%
\cite{Chapman,Mlynek,Clauser,Shimizu} for applications in atom interferometry 
and atom optics. While these elements have a number of promising features, 
integrated 
atom optical setups based on these structures still have to be developed. 
As an alternative approach, the trapping and guiding of neutral atoms
in microfabricated {\bf charged and current carrying} structures has been pursued
in recent years 
\cite{Weinstein,Hindsreview,Schmiedmayer,Haensch,Cornell,Prentiss,Engels,Hinds}. 
The simplicity and stability
of these setups as well as the favorable scaling laws 
make them promising candidates for the miniaturization and integration 
of atom optical elements.

In this paper we introduce a new approach to generate
miniaturized and integrated atom optical systems:
We propose the application of
microfabricated {\bf optical} elements
(microoptical elements)
for the manipulation of atoms and atomic matter waves
with laser light.
This enables one to exploit the vast industrial and
research interest in the field of applied optics directed towards the 
development of micro-optical elements, which has already lead to a wide range
of state-of-the-art optical system applications \cite{Herzig,Sinzinger}
in this field. Applying these elements to the field of
atom optics, however, constitutes a novel approach. 
Together with systems based on miniaturized and microfabricated 
mechanical as well as electrostatic and magnetic
devices, the application of microoptical systems 
will launch a new field in atom optics which 
we call {\bf ATOMICS} for {\bf AT}om {\bf O}ptics with {\bf MIC}ro-{\bf S}tructures.
This field will combine the unique features of devices based on the  
quantum mechanical behavior of atomic matter waves with the 
tremendous potential of micro- and nanofabrication technology.

\section {Microoptical Elements for Atom Optics}

A special attraction of using microoptical elements 
lies in the fact, that 
most of the currently used techniques in atom optics as well as laser 
cooling are based on the optical manipulation of atoms.
The use of microfabricated optical elements is therefore in many ways the canonical extension 
of the conventional optical methods into the micro-regime, so that much of the knowledge and 
experience that has been acquired in atom optics can be applied to this new regime in a 
very straightforward way. 
There are however, as we will show in the 
following, a number of additional inherent advantages in using microoptics which 
significantly enhance the applicability of atom optics and will lead to a range of new 
developments that were not achievable until now:
The use of state-of-the-art lithographic manufacturing techniques adapted from semiconductor 
processing enables the optical engineer to fabricate structures with dimensions in the 
micrometer range and submicrometer features 
with a large amount of flexibility and in a large variety of materials (glass, quartz, 
semiconductor materials, 
plastics, etc.). The flexibility of the manufacturing process allows the realization of complex 
optical elements which create light fields not achievable with standard optical components. 
Another advantage lies in the fact, that microoptics is
often produced with many identical elements fabricated in parallel on the same 
substrate, so that multiple realizations of a single conventional setup can be created
in a straightforward way. 
A further attraction of
the flexibility in 
the design and manufacturing process of microoptical components 
results from the huge potential for
integration of different elements on 
a single substrate, or, by using bonding techniques, for the integration of differently manufactured
parts into one system.
No additional restrictions arise from the small size of microoptical components 
since for most applications in atom optics, the defining parameter of an optical 
system is its numerical aperture, which for microoptical components can 
easily be as high as NA=0.5, 
due to the small focal lengths achievable.  

Among the plethora of microoptical elements that can be used for 
atom optical applications 
are refractive or diffractive microoptics, computer 
generated holograms,
microprisms and micromirrors, integrated waveguide optics,
near-field optics, and integrated 
techniques such as planar optics or micro-opto-electro-mechanical systems (MOEMS). 
Excellent overviews of microoptics can be found in  \cite{Herzig,Sinzinger}.
To our knowledge, of all these elements only computer generated 
holograms and phase gratings have been used in atom optics so far for guiding 
\cite{Schiffer} and trapping 
\cite{Michaud,Fournier,Ozeri} of 
atoms. 

In this paper we give an overview of the novel possibilities arising for atom optics
with the use of microfabricated optical elements. 
We show how all crucial components for miniaturized systems for atom 
optics, atom interferometry, and quantum information processing with neutral atoms can be 
realized with microoptical elements. We will present the key properties as well as 
the achievable parameter range  
of selected atom optical elements based on microoptical components
and will discuss the advantages and 
new possibilities that 
arise with this novel approach.

\section {Optical Dipole Force and Photon Scattering}

The optical manipulation of neutral atoms in most cases is based on the 
electric dipole interaction of atoms with laser light. It leads to 
spontaneous 
scattering of photons, which allows cooling,
state preparation and detection of atoms and to an energy
shift experienced by the atoms, which gives rise to the dipole potential.

For an understanding of the basic properties of both effects it is sufficient
to assume the atom to act as a two-level system ignoring the
details of its internal substructure \cite{Footnote1}.
A detailed treatment of 
the dipole force and the modifications arising for multi-level atoms can be found 
in \cite{Grimm,Deutsch}. 

The rate of spontaneous scattering processes is given by 

\begin{equation}
\Gamma_{sc}({\bf r})=\frac{3\pi c^2}{2\hbar\omega_0^3}\biggl(\frac{\omega_L}{\omega_0}\biggr)^3
\biggl(\frac{\Gamma}{\omega_0-\omega_L}+\frac{\Gamma}{\omega_0+\omega_L}\biggr)^2 I({\bf r}),
\label{sc_rate}
\end{equation}

valid for negligible saturation ($\Gamma_{sc} \ll \Gamma$) and large detuning
$\vert \Delta \vert \equiv \vert \omega_0 - \omega_L \vert \gg \Gamma$.
$I(\bf r)$ is the position dependent laser intensity, 
$\omega_L$ and $\omega_{0}$ are the laser frequency and the atomic resonance frequency 
respectively, 
and $\Gamma$ is the natural decay rate of the population in the excited state. 

A conservative, non-dissipative force acting on the atoms 
is derivable from the dipole potential
 
\begin{equation}
U({\bf r})=-\frac{3\pi c^2}{2 \omega_0^3}
\biggl(\frac{\Gamma}{\omega_0-\omega_L}+\frac{\Gamma}{\omega_0+\omega_L}\biggr) I({\bf r}),
\label{potential}                                                                       
\end{equation}

again valid for ($\Gamma_{sc} \ll \Gamma$) and $\vert \Delta \vert \gg \Gamma$.
The direction of the dipole force depends on the sign of the detuning 
$\Delta$.
The dipole force 
is attractive if the frequency of the laser light lies below an atomic resonance 
($\Delta<0$, "red 
detuning"), and repulsive if the frequency of the light lies above an atomic 
resonance ($\Delta>0$, "blue detuning"). 
For typical experimental conditions, the detuning is much smaller than the 
atomic resonance frequency ($|\Delta| \ll \omega_0$).
In this regime the
dipole potential scales as $I/\Delta$, whereas the rate of spontaneous scattering scales 
as $I/\Delta^2$. If decoherence as caused by spontaneous scattering has to be suppressed
the detuning should be chosen as large as possible.

It is the basic principle of the optical manipulation of atoms to create light fields
with the appropriate intensity distribution $I(\bf r)$
at a suitable detuning. Microoptical systems are extremely well suited for this purpose
since they allow the efficient and flexible generation of complex intensity 
distributions.

\section {Multiple Atom Traps}

Among the key elements in atom optics are traps for neutral atoms. Already very early
in the development of this field, a simple atom trap based on the dipole potential of 
a focused red-detuned laser beam has been realized \cite{Chu_dipole} 
and has remained an 
important element 
ever since \cite{Grimm}.
This work has been extended to the generation of multiple dipole traps by the interference 
of multiple laser beams \cite{Michaud,JESSENMEACHERGUIDONI}.

\begin{figure}
   \begin{center}
   \parbox{7.5cm}{
   \epsfxsize 7.5cm
   \epsfbox{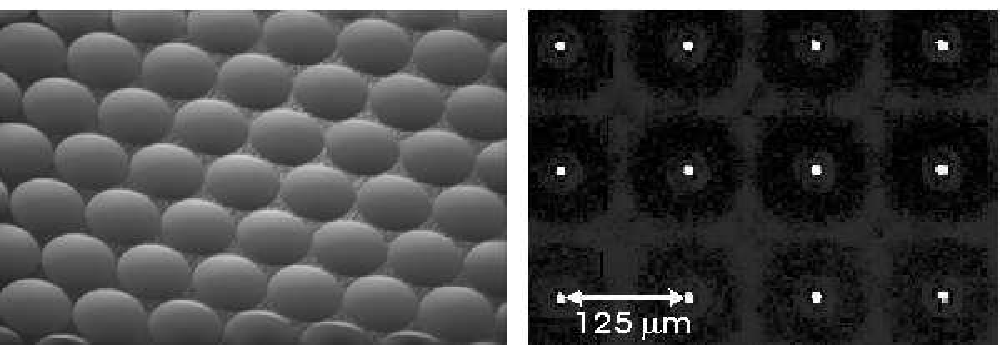}
}
   \end{center}
   \caption{Left: Hexagonal array of spherical microlenses; Right: Section of an image of the 
   intensity distribution 
   in the focal plane of a rectangular spherical microlens array (f=0.625mm, lens diameter and 
   separation:
   D=125$\mu$m).}        
   \label{fig_foci}
\end{figure}

A new approach arises from the application of one- or two-dimensional arrays of spherical 
microlenses for atom trapping (Fig. \ref{fig_foci}).    
Microlenses have
typical diameters of ten to several hundreds
of $\mu$m. Due to their short focal lengths  of typically 100$\mu$m to 1mm,
their numerical aperture can be easily as high as 0.5, resulting in foci 
whose focal size q (defined as the radius of the first minimum of the Bessel function
which results from the illumination of an individual microlens with a plane wave 
\cite{footnote_on_focal_size}) can be easily as low as q=1$\mu$m for visible laser light
\cite{Hessler}.

By focusing a single red-detuned 
laser beam with a spherical microlens array, one- or two-dimensional 
arrays of                                        
a large number of dipole traps
can be created, in which single- or multiple-atom samples can be stored 
(Fig. \ref{fig_microlenses}).

Table \ref{tabml} shows the properties of such dipole traps ($q = 1 \mu$m) for  
rubidium atoms, as an example of a frequently used atomic species, for
various commonly used laser sources \cite{footnote_on_focal_size}. 
One can easily obtain an
individual atom traps in a very 
compact setup with extremely low laser power or a
large number (100 in the case of Table \ref{tabml}) of atom traps
of considerable depth with rather moderate laser power. 
The trap depth is significantly larger than the kinetic energy of the atoms achievable with 
Doppler cooling (0.141 mK $\times k_B$ for rubidium). 
The low rates 
of spontaneous scattering that are achievable  with  sufficiently far-detuned trapping light 
ensure long storage and coherence times, while the 
strong 
localization of the atoms in the Lamb-Dicke-regime \cite{Dicke} ($x_{r}, x_{z} \ll \lambda_{0}$)
strongly suppresses heating of the atoms and makes it possible to cool the atoms to 
the ground state of the dipole potential via 
sideband cooling in all dimensions. 
In this case the size of the atomic wavefunction reaches values that are significantly smaller
than 100 nm, even approaching 10 nm in many cases,
thus making microlens
arrays well suited for the generation of strongly confined 
and well localized atom samples.

The advantages of trap arrays can, for example, be exploited for atom-interferometrical 
applications where it becomes possible to simultaneously perform
a large number of measurements (thereby improving the signal-to-noise ratio) or to 
instantaneously measure the spatial variation of the property under investigation. 
%
  
The lateral distances between the individual traps (typically 100 $\mu$m) 
make it easy to selectively detect and address the atom samples in each dipole trap.
While the natural way of addressing an individual trap
consists in sending the addressing laser beams through the corresponding microlens, 
there are also more sophisticated methods possible, e.g. with a two-photon Raman-excitation
technique as depicted in Fig. 
\ref{fig_microlenses}. Raman excitation has been applied frequently to create superposition 
states in alkali atoms \cite{Kasevich_Raman} and relies on the simultaneous interaction of the 
atoms with two mutually coherent laser fields.
For a sufficiently large detuning from the single photon resonance, only the atoms in the trap 
that is addressed by both laser beams are affected by them. By sending one Raman beam along a 
column and the other beam along a row of the lens array a specific superposition can be created 
for each individual trap, while unwanted energy shifts affecting the remaining traps can be 
easily compensated. 

These factors open the possibility to prepare and modify quantum states 
in a controlled way ("quantum engineering") in each trap, which is a necessary ingredient for 
parallelized atom interferometers and atom clocks but also for quantum 
computing, thus making 
this system also very attractive for quantum information processing applications (see section VII.
below).

\begin{figure}
   \begin{center}
   \parbox{5cm}{                                                                 
   \epsfxsize 5cm
   \epsfbox{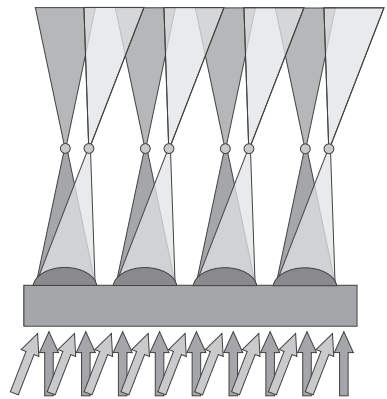}
}
   \end{center}
   \caption{Two separate dipole trap arrays created by illuminating a
   microlens array with two beams under a slightly different angle. The distance between the 
   two atom trap arrays can be altered by changing 
   the angle beween the two beams. Atoms can be transported 
   and the strength of atom-atom interactions can be controlled by changing the trap separation.}        
   \label{doublemicrolens}
\end{figure}

The manipulation of atoms with microlens arrays is extremely flexible:
It is easily possible to temporarily modify the distances between individual 
traps
if smaller or adjustable distances between traps are required. This can 
be accomplished either by using two independent microlens arrays which are laterally shifted with 
respect to each other or by   
illuminating a microlens array with two beams (possibly of different wavelength) under  
slightly different angles (Fig. \ref{doublemicrolens}), thereby
generating two distinct sets of
dipole trap arrays. Their mutual distance can be controlled by changing 
the angle 
between the two beams. With a fast beam deflector, this
can be done in real-time during the experiment. 
%
%

\section {Atomic Waveguides}

Many of the future applications of cold atomic
ensembles rely on the development of efficient means for the transport
of atoms.
Several configurations for one-dimensional guiding of atoms have been studied
in recent years. The guiding of atoms along the dark
center of blue-detuned Laguerre-Gaussian laser beams has been achieved \cite{Schiffer,Song}. 
Hollow core 
optical fibres have been used to guide atoms
%
%
\cite{Dholakia}. 
Using magnetic fields, atoms have also been guided along current carrying wires 
\cite{Denschlag,Zimmermann}, 
above surface mounted current carrying wires 
\cite{Schmiedmayer,Haensch,Cornell,Prentiss,Engels} and
inside hollow-core fibres with current carrying wires embedded in the fibre \cite{Hinds}.

\begin{figure}
   \begin{center}
   \parbox{7.5cm}{                                                                 
   \epsfxsize 7.5cm
   \epsfbox{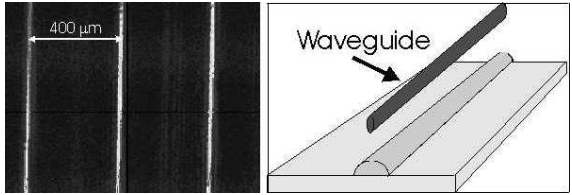}
}
   \end{center}
   \caption{Left: Section of an image of the intensity distribution 
   in the focal plane of a cylindrical microlens array (f=2.21mm, lateral lens size and separation: 
   400$\mu$m); Right: Atomic waveguide created by focusing a red-detuned
   laser beam with a cylindrical microlens.}        
   \label{cyl-foci}
\end{figure}

For atom guiding too, novel approaches arise from the application of microfabricated optical 
elements. By using cylindrical microlenses or microlens arrays, for example, 
one-dimensional guiding structures
for atoms can be developed (Fig. \ref{cyl-foci}). The light that is sent through such a system 
forms a single line-focus or a series of parallel line-foci above the lens system.

\begin{figure}
   \begin{center}
   \parbox{7.5cm}{                                                                 
   \epsfxsize 7.5cm
   \epsfbox{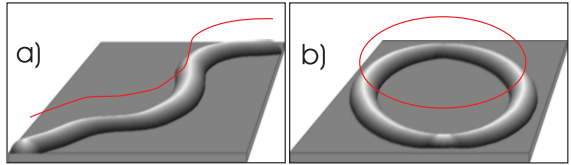}
}
   \end{center}
   \caption{Complex waveguide geometries achievable with microfabricated lenses. 
   (a) Curved guiding structure; (b) Storage ring or closed-loop waveguide acting as
   resonator for atomic matter waves.}        
   \label{complicated_cylindrical_microlenses}
\end{figure}

By focusing a red-detuned laser beam with homogeneous intensity distribution an atomic 
waveguide is formed. Atoms are confined in the two dimensions perpendicular to the lens 
axis but are free to propagate along 
the axis. 
Since the manufacturing process is identical to that of spherical microlens arrays,
cylindrical microlenses can be manufactured with 
similar dimensions and numerical apertures so that the focal size of the line foci 
again can be as low as 
q=1 $\mu$m for visible light.
Due to the flexibility of the fabrication process, rather complicated waveguide geometries 
can be achieved. The shape of the lens can be curved 
(Fig. \ref{complicated_cylindrical_microlenses} (a)) creating a bent waveguide or the lens can 
even be circular (Fig. \ref{complicated_cylindrical_microlenses} (b)) resulting in a closed 
one-dimensional potential minimum which can be used as a miniaturized storage ring 
or resonator for atomic 
matter waves.

Table \ref{tabcyl} shows the properties of a 10mm long waveguide for $q = 1 \mu$m,
rubidium atoms
and various 
commonly used laser sources.
Waveguides with potential depths comparable to the kinetic energy achievable with
Doppler cooling
can easily be created.
They are well suited for generating single or integrated structures for guiding
atoms and atomic matter waves in compact atom optical systems although other guiding 
structures 
\cite{Schmiedmayer,Haensch,Cornell,Prentiss,Engels,Hinds,Schiffer,Song,Dholakia,Denschlag,Zimmermann}
might be more favorable 
for guiding atoms over distances that are larger than several cm. 

Of specific interest for atom optical applications are single-mode waveguides for atomic 
matter
waves since the development of atom interferometers based on guided atoms and the 
coherent transport of Bose-Einstein condendates or atom laser outputs are strongly profiting
from single-mode guiding. For this purpose it is essential to have the atoms confined in 
the Lamb-Dicke regime.
%
%
As can be seen from Table \ref{tabcyl}, with the exception of the 
$CO_2$ waveguide, the oscillation frequencies are 
comparable to (in the z-direction) or significantly larger than (in the r-direction) the 
recoil frequency $\omega_{R}$ (= $24 \times 10^3 s^{-1}$ for rubidium). 
Thus, single-mode guiding with sufficiently low probability of excitation of the 
atomic wave packets to higher vibrational states is achievable with very low rates
for spontaneous scattering.

In addition to microfabricated cylindrical lenses, there exist other microoptical techniques 
which 
can be applied to create waveguide structures for atoms. One of the most promising candidates
for large scale integration of atomic waveguides is the application of 
planar optical waveguide structures \cite{Herzig,Sinzinger}.
The basic concept is to implement spatial variations in the index of refraction near the
surface of an 
optical substrate
in order to induce total internal reflection at the boundaries of these regions and thus 
spatially localize the light field. 
The structured light field evanesces from the surface and can be used to create complex
two-dimensional patterns of dipole potential wells and
barriers 
parallel to the
surface of the substrate. 
Additional confinement of the atoms in the direction perpendicular to the surface 
can be achieved for example by a 
standing light wave pattern created by reflecting a laser beam from the surface 
\cite{Gauck} or
by an additional magnetic field creating a two-dimensional magnetic waveguide \cite{Hindswaveguide}.

\section {Beam Splitters and Interferometers}

Significant applications of microfabricated atom-optical and
atom-interferometrical devices arise when the key elements for more complex 
systems can be demonstrated and their integration
can be realized. One of the key elements
in this respect is a beam splitter for atomic matter waves. 
Beam splitters based on microfabricated current carrying wires have already been demonstrated
\cite{muller,casse}.
Microoptical beam splitters on the other hand, can be based on a straightforward 
extension
of the microoptical waveguides discussed above. The demonstration of coherent beam splitting 
will be the next step for all of these devices (see also \cite{Sanpera}).

\begin{figure}
   \begin{center}
   \parbox{7.5cm}{                                                                 
   \epsfxsize 7.5cm                       
   \epsfbox{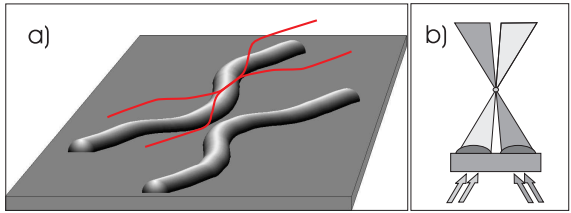}
}
   \end{center}
   \caption{(a) Beam splitter for atomic matter waves based on two microlens waveguides. 
   (b) The microlenses are illuminated under different angles in order to make
   the laser foci overlap at the center of the beam splitter. Atomic wave 
   packets entering along one waveguide are split into up to four output wave packets.}        
   \label{beamsplitter}
\end{figure}

Fig. \ref{beamsplitter} shows a beam splitter based on the combined light fields
of two curved cylindrical microlenses.
Each microlens
is illuminated by a light beam under a different 
angle (Fig. \ref{beamsplitter} (b)). This causes the laser foci to be laterally 
displaced with respect to the center of the lenses. By an appropriate
choice of the displacements, the guiding potentials of the waveguides can be made to overlap
at the closest approach making the full structure to act as a beam splitter. 
With an appropriate choice of the polarization states of the two light fields, interference 
effects can be avoided.
%
%
Input wave packets propagating
along one of the waveguides are split into up to four output wave packets propagating 
along both waveguides in both directions. 
Since the full structure is completely based on conservative potentials,
the beam splitting process should be non-dissipative, so that coherent beam
splitting of atomic matter waves is achievable. A detailed discussion 
of the properties of this beam splitter is published elsewhere \cite{Sanpera}.
%

\begin{figure}
   \begin{center}
   \parbox{7.5cm}{                                                                 
   \epsfxsize 7.5cm
   \epsfbox{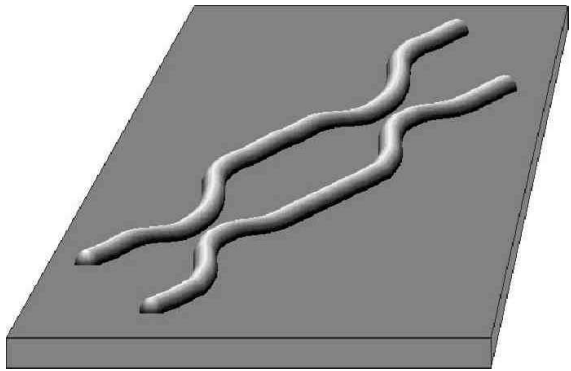}
}
   \end{center}
   \caption{Mach-Zehnder type interferometer based on the combination of two 
   beam splitters of Fig. \ref{beamsplitter}.
   The enclosed area between the beam splitters can be as large as 1$cm^2$ for
   typical laser powers.}        
   \label{sagnac}
\end{figure}

This beam splitter can easily be extended to a microfabricated
interferometer for atomic matter waves: Combining two beam splitters
creates a Mach-Zehnder-type interferometer (Fig. \ref{sagnac}). The two wave packets propagating along
the output waveguides of the first beam splitter
can be recombined in a second beam splitter and made to interfere.

As shown in section V, with typical laser powers 
guiding of ultracold atomic matter waves along microlens-based waveguides 
over distances of several 10 mm
is possible. This allows the realization of
interferometers with an enclosed area $A$ of at least 1 $cm^2$
which represents the state-of-the-art for atomic matter wave interferometers \cite{Gustavson}. 
With microfabrication such a large area can be achieved in a system with dimensions that are
significantly smaller than those of setups based on conventional methods.
This presents a major step towards miniaturization of
atom-interferometrical devices and promotes the wide applicability of sensors
based on atom interferometry. The Mach-Zehnder interferometer
presented in Fig. \ref{sagnac} can be used as a Sagnac interferometer
for measuring rotational motion with extremely high sensitivity. 
%

\section {Quantum Information Processing}

\begin{figure}
   \begin{center}
   \parbox{7.5cm}{
   \epsfxsize 7.5cm
   \epsfbox{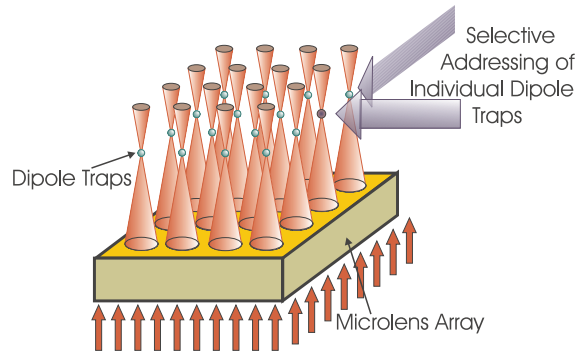}
}
   \end{center}
   
   \caption{Two-dimensional array of dipole traps created by focusing a red-detuned laser 
   beam with an array of microlenses. Due to their large separation (typically 100$\mu$m)
   individual traps can be addressed selectively, e.g. by two-photon Raman-excitation, 
   as depicted.}        
   \label{fig_microlenses}
\end{figure}

The microoptical elements discussed so far lend themselves also particularly well to quantum  
information processing \cite{Bouwmeester} 
exploiting the fact that qubits can be represented by superpositions of internal and external
states of atoms in a straightforward fashion.
%
%
Inherently an approach based on microoptical systems 
addresses two of the most important requirements for the
technological implementation of quantum information processing: parallelization and scalability.
In addition, the possibility to selectively address individual qubits 
is essential for most schemes proposed for quantum 
computing with 
neutral atoms. 
As a first and easily achievable implementation, 
single qubits associated with long-lived internal states 
can be prepared and rotated in each 
individual trap of a two-dimensional dipole trap array (Fig. \ref{fig_microlenses}), 
stored and later read out again. Thus, this device can serve as a quantum state register.
%
%
The low rates of spontaneous scattering that are achievable in far-detuned
microtraps (see Table \ref{tabml}) ensure the long coherence times that are required for 
succesful quantum information processing. 

An important step towards a functioning quantum computer is the implementation of two-qubit 
gates.
Due to their inherent features, microoptical devices are also 
well suited for
this purpose. 
Considering, for example, quantum gates based on dipole-dipole interactions between atoms 
\cite{Brennen} all requirements are fulfilled in the configuration
depicted in Fig. \ref{doublemicrolens}. Atoms localized in neighboring traps can be
brought close to each other with a definable separation in the single-micron range 
and for a predefined duration, in order 
to collect the required phase shift. Especially well suited is this configuration also for 
quantum gates based on the 
dipole-dipole interaction of low-lying Rydberg states in constant electric fields, as proposed 
in \cite{Jaksch2}.
   
\begin{figure}
   \begin{center}
   \parbox{7.5cm}{                                                                 
   \epsfxsize 7.5cm
   \epsfbox{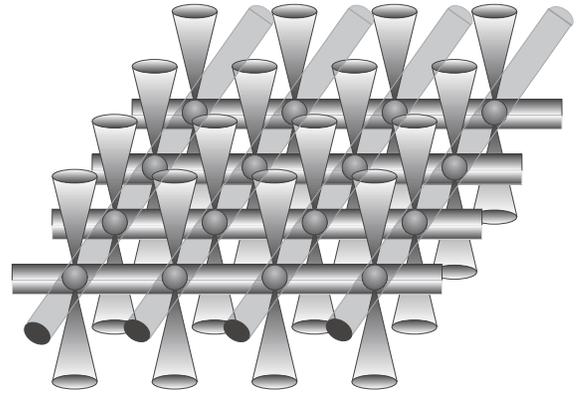}
} 
   \end{center}
   \caption{Quantum information network. Superimposing the foci of a spherical microlens array 
   with those 
   of a cylindrical microlens array results  in a network of interconnected dipole traps.
   Atoms can be transported between the traps along the atomic waveguides
   and entanglement via controlled collisions can be induced.}        
   \label{Network}
\end{figure}

Another possible implementation of quantum gates relies on the entanglement of atoms via 
controlled cold collisions \cite{Jaksch1} which can be realized in configurations that consist
of a combination of spherical and cylindrical microlenses. As depicted in Fig. \ref{Network}, 
the two-dimensional array of dipole traps shown in Fig. \ref{fig_microlenses} can be combined
with waveguides created by cylindrical lens arrays. The result is a network of 
interconnected dipole traps. This network can be utilized for the transport of atoms from one 
potential well to the other by applying additional laser fields. Since this transfer can be 
made state dependent - via the detuning of the transport light fields - this mechanism can 
be utilized 
for letting the atoms collide state-dependently, thereby allowing the implementation of
controlled-collision quantum gates.

\begin{figure}
   \begin{center}
   \parbox{7.5cm}{                                                                 
   \epsfxsize 7.5cm
   \epsfbox{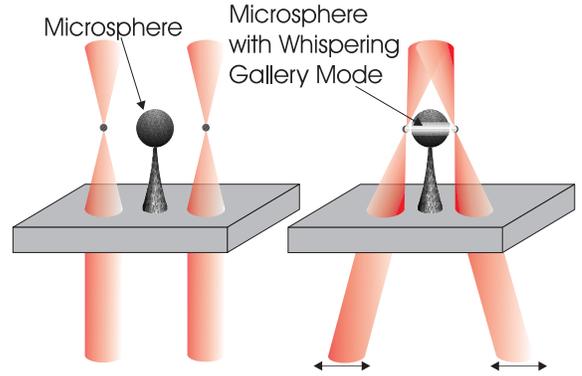}
}
   \end{center}
   \caption{Entanglement of atoms in optical microtraps via their interaction
   with the high-Q mode of a microsphere resonator. By moving the trapping
   potential (see Fig. \ref{doublemicrolens}) atoms can be brought into contact with the 
   resonator mode in a controllable fashion. This system allows the implementation
   of fast quantum gates for quantum computing.}        
   \label{microsphere}
\end{figure}

A further, very powerful method for two- or multiple-atom entanglement is based on the extremely
strong coupling between atoms and single-mode light fields exploited in cavity-QED experiments
\cite{Berman,Special}.
Fig. \ref{microsphere} shows a configuration where two atoms trapped in microoptical dipole traps
based on spherical microlenses (Fig. \ref{fig_microlenses})
can interact with the whispering-gallery mode of a dielectric microsphere 
\cite{Braginsky,Collot,Mabuchi,Gorodetsky}. 
The interaction with this mode, which can 
have a quality factor Q as high as $10^{10}$ represents an extremely fast method of atom entanglement. 
%

\begin{figure}
   \begin{center}
   \parbox{7.5cm}{                                                                 
   \epsfxsize 7.5cm
   \epsfbox{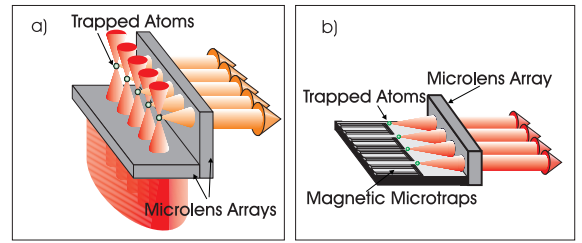}
}
   \end{center}
   \caption{Spatially resolved readout of the internal and external states
   of atoms (e.g. the state of a qubit) using microlens arrays: (a) Integration
   of two spherical microlens arrays creates a combined system of dipole
   traps and efficient detection optics.
   (b) Integration of a microlens array (for readout) with microfabricated 
   magnetic or electrostatic trapping
   structures.}        
   \label{readout}
\end{figure}

Due to their large numerical aperture, microoptical components can also be used
for efficient spatially resolved read-out of quantum information of atoms 
(Fig. \ref{readout}). 
In most cases the state of a qubit is recorded by exciting the atom state-selectively with 
resonant light and collecting the fluorescence light. Microoptical components can be used for 
the collection optics.
%
%
Furthermore, the optical detection of quantum states with microoptical components is
not restricted to optical trapping structures (Fig. \ref{readout} (a)) but can also be combined 
with the miniaturized 
magnetic and electric trapping
structures (Fig. \ref{readout} (b)) currently investigated by a number of groups 
\cite{Schmiedmayer,Haensch,Cornell,Prentiss,Engels,Hinds}.

Thus, all steps required for quantum information processing with neutral atoms -
i.e. the preparation, manipulation and storage of qubits, entanglement, and the efficient 
read-out of quantum information - 
can be performed using microfabricated optical elements.

\section {Microstructured Atom Sources}

An important step towards fully integrated atom optical setups is the development of 
miniaturized
sources of ultracold atoms. Again, microoptical components can be used to achieve this goal,
once more profiting from the fact that most of the preparation techniques for atom samples
are based on optical manipulation.
In this section we describe a miniaturized version of the "workhorse" in atom optics, the 
magneto-optical trap (MOT). Our approach for developing a compact MOT 
is based on the concept of "planar optics" 
\cite{Herzig,Sinzinger}.
%

In planar optics, complex optical systems are integrated monolithically on a
substrate.
The optical path is folded in such a way, that the light propagates 
along a zig-zag path inside the substrate 
and 
the light is 
%
%
manipulated by reflective optical components (mirrors, beam splitters, lenses, 
retarders, etc.) that are mounted on or machined into the surface of 
the substrate. 
%
Planar optical systems can be used for a broad range of atom optical applications including 
atom traps, waveguides, beam splitters, interferometers, networks, and systems for 
quantum information processing. 

\begin{figure}
   \begin{center}
   \parbox{7.5cm}{                                                                 
   \epsfxsize 7.5cm
   \epsfbox{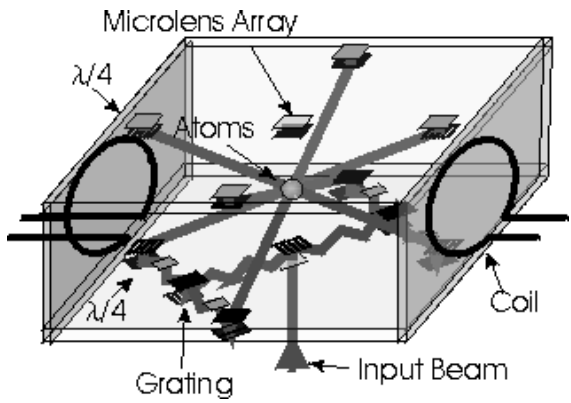}
}
   \end{center}
   \caption{Integrated magneto-optical trap (MOT) based on planar optics. The configuration
   consists of two optical substrates mounted in parallel planes and a pair of 
   quadrupole coils. All optical elements needed for the operation of a MOT are mounted on 
   or machined into the two substrates (see text).
   The substrates can also contain additional microoptical elements, e.g. a 
   microlens array, as shown here. 
   }        
   \label{planaropticsMOT}
\end{figure}

As one example of an integrated system based on planar optics,  
Fig. \ref{planaropticsMOT} depicts an integrated MOT.
The optical components are mounted on 
two parallel substrates, separated by several mm or cm. The trapping light is coupled into
the lower substrate and split into four beams, which, after passing through quarter-wave plates,
intersect a few mm or cm above the substrate. The beams enter the upper substrate and are 
retroreflected after double-passes through quarter-wave plates. Between the two substrates 
microfabricated coils for the MOT quadrupole fields are mounted. At the center of the upper 
substrate a two-dimensional microlens array is included indicating the possibility of 
integrating the MOT with other microoptical elements. The small size of 
an integrated MOT configuration will limit the achievable atom number.
With a beam diameter of several mm still 
several $10^6$ atoms should be trapped imposing no limitations for most of the applications 
discussed here.
%

\section {Integration}

The huge potential for integration of microoptical components 
can be used for a large variety of further atom optical purposes. Since the same techniques are 
applied for the fabrication of microoptical components and microstructured wires on surfaces,
microoptical components can be easily combined with the magnetic and electric strucures 
of \cite{Weinstein,Hindsreview,Schmiedmayer,Haensch,Cornell,Prentiss,Engels,Hinds} 
(e.g., as shown Fig.\ref{readout} (b)).
In addition, microfabricated atom-optical components can be integrated with optical 
fibres and waveguides
so that the results of atom optical operations after being read-out 
by detection of the scattered light 
can be further processed by optical means.
Another canonical extension is given by the integration of microoptical components with 
optoelectronic devices such as semiconductor laser sources and photodiode detectors. 
In this case, the 
communication with the outside world can take place fully electronically,
with the required laser light created in situ and the optical signals converted back to 
electrical signals on the same integrated structure.
%
     
The richness of the available microfabrication technologies manifests itself in another 
approach to integration based on micro-opto-electro-mechanical 
systems (MOEMS) \cite{Herzig,Sinzinger,Motamedi}. The defining purpose of these elements is to 
steer optical signals by microfabricated mechanical components which can be positioned 
electrically.
Significant research effort has been directed towards the development of microfabricated 
opto-electro-mechanical
components. 
%
%
With these elements fast switching and steering of laser 
beams becomes
possible, so that complex spatial and temporal
field patterns can be generated. 
%

\section {Experimental Considerations}

One important question 
that has to be addressed in view of the applicability of microoptical systems is how 
to load atoms into these microscopic traps and guides. 
%
%
Since the distance of the substrate carrying these structures
to the atom sample for microoptical components is typically given by the focal length of 
several hundred $\mu$m, loading of atoms should be easier than for 
microstructured systems based on the magnetic or 
electrostatic fields
\cite{Weinstein,Hindsreview,Schmiedmayer,Haensch,Cornell,Prentiss,Engels,Hinds}
with distances on the order of 1 to 100 $\mu$m. 
Loading into these magnetic microstructures has been achieved by 
modified MOT configurations and a transfer of the atoms to the substrate's
surface
%
\cite{Schmiedmayer,Haensch} or by elaborate schemes to transfer the atoms to the surface
\cite{Cornell,Prentiss,Engels,Hinds}. 
Similar solutions can be applied for systems based on microoptical elements
since for small microoptical structures the laser beams for a MOT 
can be reflected off or
transmitted by the structure carrying substrate with only minor perturbations of the
MOT. 

In addition, further
loading techniques are available for optical microstructures: 
1) Microoptical systems can be positioned several mm or even cm away from the MOT with the focal 
plane 
being imaged into the center of the MOT without significant degredation
thus significantly enlarging the distance between the MOT and the structure carrying substrate. 
2) The compact size 
of microoptical elements together with the transmittivity of the supporting substrate 
should enable one to position the substrate directly at the required distance from the 
MOT by simply shining parts of the MOT beams through the transparent substrate, 
thus allowing to capture atoms 
in the micropotentials in situ or, if necessary, after a small lateral translation at the 
end of the MOT phase.

In the long term, the canonical 
way to load atoms into these devices will be based on microstructured atom sources as 
described in section VIII.,
or on the transfer of atoms collected in a separated MOT and then transferred into
the microfabricated atom-optical systems by guiding them with laser beams, magnetic fields 
or inside guiding structures based on some of the systems discussed above.
 
Another point that has to be considered, 
which is also present for magnetic and electrostatic structures, are the possible side-effects of 
stray fields. In the case of microoptics, especially for more complex and integrated 
devices, unwanted scattering, diffraction, 
and reflection of the light fields has to be suppressed. 
Due to the small distance of the atoms from the structure carrying substrates these problems might be
more severe than for conventional optical components. 
Measures have to be taken to minimize these side effects and 
to absorb or guide the light fields 
away from the atoms after their desired interaction. Techniques for this purpose (e.g., 
beam-dumping)
are, however, well established in standard optics and can also be applied for the structures 
discussed here. In addition, similar questions arise for standard applications of microoptical
components in applied optics and the design of the components is optimized to reduce these effects.

For all of the configurations presented above, fluctuations of the depth, shape, and position 
of the trapping potentials have to be sufficiantly small. This is especially true for
applications requiring a preservation of coherence. The effects of beam jitter, intensity fluctuations, 
etc. have to be
investigated and minimized if necessary although integrated microoptical systems
have the advantage of being instrinsically stable.

Finally, the availability of microoptical components has to be discussed. Standard components 
such as arrays of spherical or cylindrical microlenses are commercially available. Optimized 
or highly integrated components can be custom-made with available micro- and nanofabrication
technology by several companies.
An excellent overview of the state-of-the-art of microoptics design and manufacturing can be found 
in references \cite{Herzig,Sinzinger}.

\section {Conclusion}

In this paper we have shown that by using microoptical components one can create 
a variety of powerful configurations with many applications in the fields of atom optics, atom 
interferometry, and quantum information processing with neutral atoms. These applications 
hugely benefit from the many inherent advantages of microoptical 
components.
\newline
- {\bf Importance of optical methods:} In atom optics, atom 
interferometry and quantum information processing with neutral atoms, the initial preparation
of atom samples, the preparation and 
manipulation of superposition states and qubits and the readout of results are usually 
achieved by optical 
means. 
For this reason the application of state-of-the-art microoptical systems for the optical 
manipulation of atoms is the canonical extension of today's experimental techniques into the 
microregime.

Another potential advantage 
of the application of microoptical systems relates to 
the fact that in the microscopic magnetic and electric structures of 
\cite{Weinstein,Hindsreview,Schmiedmayer,Haensch,Cornell,Prentiss,Engels,Hinds} 
the atoms have to be very close 
(typically 1-100$\mu$m) to a room-temperature metallic surface in order to ensure the large
potential depths required for the envisaged applications. As calculated in 
\cite{Henkel}, this can significantly reduce the trapping and coherence times due to the 
coupling of the atoms to fluctuating magnetic fields in the vicinity of the surface. 
In the case of
microoptical systems this problem is insignificant, since for glass substrates the 
effects of magnetic fluctuations are significantly reduced, and since the distance of the atoms 
to the surface is 
much larger (typically several hundreds of $\mu$m) for the same potential depths. 
Optical trapping potentials on the other hand, intrinsically suffer from possible losses of 
coherence due to spontaneous scattering of photons. While this effect can in principle
be suppressed by employing light fields with sufficiently large detuning 
the final limit of the trapping and coherence 
times in optical traps remains to be investigated.    
%
%
\newline
- {\bf Size:} The small size of microoptical components ensures a decrease in volume, weight
and, ultimately, cost of the setups. As shown, this does not negatively 
affect the attainable optical properties since the numerical aperture of microoptical components 
can be very high.   
\newline
- {\bf Design flexibility:} Microoptical components are fabricated using lithographic 
fabrication techniques that are adapted from semiconductor processing. The use of these 
techniques allows for a large amount of flexibility in the design of microoptical 
components. Devices that are impractical or impossible to fabricate under the constraints of 
the 
conventional fabrication techniques for optical components, can now easily be designed and 
fabricated, thereby opening new possibilities for optical atom manipulation. 
\newline
- {\bf Scalability:} 
Microoptical systems 
can also be easily produced with many identical elements fabricated in parallel on the same substrate,
so that multiple realizations of atom optical configurations are easily achievable without
the unwanted side effect of loosing individual addressability. Thus, massive parallelization
and the development of extremely complex systems can be achieved.
\newline
- {\bf Integrability:} The utilization of the same fabrication techniques also enables one to 
combine microoptical 
components with optoelectronic devices.
%
For the purpose of the optical manipulation of atoms 
the integrability of microoptical components offers enormously fruitful possibilities for 
future developments. In addition to utilizing complex configurations consisting of several
microoptical components, integration with microfabricated magnetic and electric structures 
for atom manipulation but also with detectors and laser sources can 
be foreseen.

In summary, we have introduced the new research direction of using microfabricated optical 
elements for atom optics,
atom interferometry, and quantum information processing with neutral atoms. We have presented 
a variety of possible configurations with widespread applications and have discussed the 
advantages of this new approach. 
Together with the currently investigated, 
promising techniques of microfabricated mechanical, magnetic and electric structures
\cite{Chapman,Mlynek,Clauser,Shimizu,Weinstein,Hindsreview,Schmiedmayer,Haensch,Cornell,Prentiss,Engels,Hinds}, our approach
opens the possibility
to combine the principal advantages of quantum mechanical systems with the advanced 
technological basis of micro- and nanofabrication and will
lead to integrated setups, 
that will significantly enhance 
the applicabiltiy of integrated atom optics.

\section {Acknowledgements}

We thank A. Sanpera, M. Lewenstein, and P. Zoller for helpful discussions on quantum 
information processing with neutral atoms.
This work is supported by the program ACQUIRE (IST-1999-11055) of the European Commission 
and the 
SFB 407 of the {\it Deutsche 
Forschungsgemeinschaft}.


%
%
%

%
%

%
%

%
%

%
%

%
%

%
%

%
%

%
%

%
%

%
%

%
%

%
%

%
%

%
%

\onecolumn
\renewcommand{\baselinestretch}{1}
\begin{table}[htb]
\centerline{\bf Table I}
{\center
\begin{tabular}{|c|c||c|rc|rc|c|}
 & Power per & Potential & \multicolumn{2}{c|}{Vibrational} 
& \multicolumn{2}{c|}{Ground State} & Scattering \\

Laser & Lenslet & Depth & \multicolumn{2}{c|}{Frequency [$s^{-1}$]} & 
 \multicolumn{2}{c|}{Size [nm]} & Rate\\
 
 & [mW] & [mK $\times$ $k_{B}$] & \hspace{2em}$\omega_{r}$
& $\omega_{z}$ & \hspace{1em}$x_{r}$ & $x_{z}$ & {$\Gamma_{sc}$ [$s^{-1}$]}\\ \hline\hline
Diode & & & & & & & \\ 

($\lambda_{L}$=783nm) & \raisebox{3ex}{1} & \raisebox{3ex}{6.1} & \raisebox{3ex}{$1.5 \times 10^6$} 
& \raisebox{3ex}{$2.0 \times 10^5$} & \raisebox{3ex}{22} 
& \raisebox{3ex}{61} & \raisebox{3ex}{3800} \\ \hline

Ti:Sapphire  & & & & & & &\\ 
($\lambda_{L}$=783nm) & \raisebox{3ex}{10} & \raisebox{3ex}{61} & \raisebox{3ex}{$4.7 \times 10^6$} 
& \raisebox{3ex}{$6.4 \times 10^5$} & \raisebox{3ex}{13} 
& \raisebox{3ex}{34} & \raisebox{3ex}{38000} \\ \hline

Ti:Sapphire  & & & & & & &\\ 
($\lambda_{L}$=830nm) & \raisebox{3ex}{10} & \raisebox{3ex}{7.8} & \raisebox{3ex}{$1.7 \times 10^6$} 
& \raisebox{3ex}{$2.3 \times 10^5$} & \raisebox{3ex}{21} 
& \raisebox{3ex}{57} & \raisebox{3ex}{270} \\ \hline

Nd:YAG  & & & & & & & \\ 
($\lambda_{L}$=1064nm) & \raisebox{3ex}{100} & \raisebox{3ex}{18} & \raisebox{3ex}{$2.5 \times 10^6$} 
& \raisebox{3ex}{$3.5 \times 10^5$} & \raisebox{3ex}{17} 
& \raisebox{3ex}{47} & \raisebox{3ex}{63} \\ \hline

$CO_{2}$ & & & & & & & \\ 
($\lambda_{L}$=10,6$\mu$m) & \raisebox{3ex}{1000} & \raisebox{3ex}{0.80} & \raisebox{3ex}{$5.4 \times 10^4$} 
& \raisebox{3ex}{$1.0 \times 10^4$} & \raisebox{3ex}{120} 
& \raisebox{3ex}{270} & \raisebox{3ex}{1.3x$10^{-3}$} \\ \hline
\end{tabular}}
\vspace{0.5cm}
\caption{Properties of dipole traps for $^{85}$Rb atoms ($\omega_{0}$ = 2$\pi$c/$\lambda_{0}$, 
with $\lambda_{0}$ = 780.0 nm, $\Gamma$ = 2$\pi \times$ 5.89 MHz) generated by a two-dimensional
microlens array (Fig. \ref{fig_microlenses}) for various commonly used laser sources.
The parameters are calculated 
for a focal size of q=1$\mu$m ($CO_{2}$-Laser:
q=10$\mu$m).
The power per lenslet is chosen in such a way 
that for typical laser sources 100 dipole 
traps can be generated simultaneously. The r-direction is the 
direction perpendicular, the z-direction is the direction parallel to the laser beam.}
\label{tabml}
\end{table}

\renewcommand{\baselinestretch}{1}
\begin{table}[htb]
\centerline{\bf Table II}
\rule{0mm}{3mm}
{\center
\begin{tabular}{|c|c||c|rc|rc|c|}
& Power & Potential & \multicolumn{2}{c|}{Vibrational} 
& \multicolumn{2}{c|}{Ground State} & Scattering \\

Laser && Depth & \multicolumn{2}{c|}{Frequency [$s^{-1}$]} & 
 \multicolumn{2}{c|}{Size [nm]} & Rate\\
 
 & [W] & [$\mu$K $\times$ $k_{B}$] & \hspace{2em}$\omega_{r}$
& $\omega_{z}$ & \hspace{1em}$x_{r}$ & $x_{z}$ & {$\Gamma_{sc}$ [$s^{-1}$]}\\ \hline\hline
Diode & & & & & & & \\ 

($\lambda_{L}$=783nm) & \raisebox{3ex}{0.1} & \raisebox{3ex}{59} & \raisebox{3ex}{$1.5 \times 10^5$} 
& \raisebox{3ex}{$2.0 \times 10^4$} & \raisebox{3ex}{72} 
& \raisebox{3ex}{190} & \raisebox{3ex}{37} \\ \hline

Ti:Sapphire  & & & & & & &\\ 
($\lambda_{L}$=783nm) & \raisebox{3ex}{1} & \raisebox{3ex}{590} & \raisebox{3ex}{$4.6 \times 10^5$} 
& \raisebox{3ex}{$6.3 \times 10^4$} & \raisebox{3ex}{40} 
& \raisebox{3ex}{110} & \raisebox{3ex}{370} \\ \hline

Ti:Sapphire  & & & & & & &\\ 
($\lambda_{L}$=830nm) & \raisebox{3ex}{1} & \raisebox{3ex}{75} & \raisebox{3ex}{$1.6 \times 10^5$} 
& \raisebox{3ex}{$2.3 \times 10^4$} & \raisebox{3ex}{67} 
& \raisebox{3ex}{180} & \raisebox{3ex}{2.6} \\ \hline

Nd:YAG  & & & & & & & \\ 
($\lambda_{L}$=1064nm) & \raisebox{3ex}{10} & \raisebox{3ex}{170} & \raisebox{3ex}{$2.5 \times 10^5$} 
& \raisebox{3ex}{$3.6 \times 10^4$} & \raisebox{3ex}{53} 
& \raisebox{3ex}{140} & \raisebox{3ex}{0.6} \\ \hline

$CO_{2}$ & & & & & & & \\ 
($\lambda_{L}$=10,6$\mu$m) & \raisebox{3ex}{100} & \raisebox{3ex}{77} & \raisebox{3ex}{$1.7 \times 10^4$} 
& \raisebox{3ex}{$3.1 \times 10^3$} & \raisebox{3ex}{210} 
& \raisebox{3ex}{490} & \raisebox{3ex}{1.3x$10^{-4}$} \\ \hline
\end{tabular}}
\vspace{0.5cm}
\caption{Properties of one-dimensional waveguides for $^{85}$Rb atoms 
based on light focused by a cylindrical microlens for various commonly used 
laser sources. The parameters are 
calculated for a length of the waveguide of 10 mm and
for a focal size of q=1$\mu$m ($CO_{2}$-Laser:
q=10$\mu$m).
}
\label{tabcyl}
\end{table}


\begin{references}
\bibitem{Footnote1} For the quantitative discussion given in Table I and II we performed
a weighted average of 
different components of the interaction assuming the light
to be linearly polarized.
\end{references}
\end{document}